\documentclass{article}
\title{Coherent states in real parameterization up to SU(5) and classical dynamics of spin systems }
\author{Kh. Kh. Muminov, Y. Yousefi  \\
 Physical-Technical Institute named after S.U.Umarov \\
  Academy of Sciences of the Republic of Tajikistan \\
  Aini Ave 299/1, Dushanbe, Tajikistan, \\
  e-mail: khikmat@inbox.ru, muminov@phti.tj}
\date{}
\begin{document}
\maketitle
\begin{abstract}
In this paper, we develop the formulation of the spin coherent state in real parameterization up to SU(5). The path integral in this representation of coherent state and its classical consequence are investigated. Using the resolution of unity of the coherent state, we derive a path integral expression for transition amplitude and in the classical limit we derive the classical equation of motion.
\end{abstract}
\section{Introduction}
Coherent states were originally constructed and developed for the Heisenberg-Weyl group to investigate quantized electromagnetic radiation [1]. These coherent states were generated by the action of the Heisenberg-Weyl group operators on the vacuum state which led to theoretical group generalization by Peleromov [2] and Gilmore [3]. These two mathematical frameworks differ in some points, such as the representations of groups and the reference states, these differences are summarized in [4].

The set of coherent states elaborated in this paper allow one to see explicitly that the reason of arising of electric quadrupole field is related to relativistic nature [5]. Antiferromagnets, due to their structure, in the states close to the vacuum (basic) one could manifest the existence of external quadrupole electric field [6,7], theoretical bases of this were provided earlier by Dzyaloshinskii [8].

In this paper, we construct a set of explicit coherent states up to SU(5), and apply theoretical group techniques to facilitate the investigation of nonlinear quantum systems and quantum entanglement. In order to construct explicit coherent states, we need to specify the group representation and the reference states. We consider the reference state as $ (1,0,..,0)^T$ , where T denotes transposition. This states is a highest weight state, in the sense that it is annihilated by each of the SU(n) raising operators.

In this paper we construct the set of generalized spin coherent states up to SU(5) group in an explicit form via real parameterization. By use of path integral technique [9] we construct the Lagrangian and equations of motion for the semiclassical dynamics of spin multipole systems for higher magnitude of spins up to 2.    

\section{Properties of SU(2) group}

For construction coherent state in SU(2), we consider the reference state as $ (1,0)^T $, the general form of coherent state in this group we obtain form the following formula [10]:

\begin{eqnarray}
| \psi \rangle = e^{-i\phi S^z}e^{-i\theta S^y}|0\rangle= C_0 |0\rangle+C_1 |1\rangle
\end{eqnarray}

that

\begin{eqnarray}
 C_0=cos(\theta/2)e^{-i\phi/2} &   & {    }  C_1=sin(\theta/2)e^{i\phi/2}
\end{eqnarray}

Here we consider classical counterparts of the spin operators and obtain their average. The vector 

\begin{eqnarray}
\vec S=\langle \psi | \vec {\hat S} | \psi  \rangle
\end{eqnarray}

can be regarded as a classical spin vector, and 

\begin{eqnarray}
Q^{ij}=\langle \psi | \hat S^i \hat S^j | \psi  \rangle
\end{eqnarray}

can be regarded as a component of quadrupole moment. Because the spin operators at different lattice sites do not commute, we have for all such products

\begin{eqnarray}
\langle \psi | \hat S_i^i \hat S_{i+1}^j | \psi  \rangle=\langle \psi | \hat S_i^i  | \psi  \rangle \langle \psi | \hat S_{i+1}^j | \psi  \rangle
\end{eqnarray}

Average spin expression for the SU(2) group is

\begin{eqnarray}
\langle S^+  \rangle &=&e^{i\phi} sin\theta \nonumber\\
\langle S^-  \rangle&=&e^{-i\phi} sin\theta \nonumber\\
\langle S^z  \rangle&=&cos\theta
\end{eqnarray}

Let us consider a Hamiltonian $\hat H $   acting in Hilbert space. We shall assume that $ \hat H$  can be expanded as the finite polynomial of the infinitesimal operators $ \hat S^\pm $   and $ \hat S^z $   of SU(2). The transition amplitude (propagator) from the state $| \psi \rangle$   at time t to the state  $| \psi_1 \rangle$   at time $ t_1$ is given by

\begin{eqnarray}
P(\psi_1,t_1;\psi,t)=\langle \psi_1 | exp(-\frac{i}{\hbar}(\hat H(t_1-t)))| \psi \rangle
\end{eqnarray}

In order to derive the path integral from the amplitude P, we divide $ (t_1-t) $ into n equal time intervals $ \epsilon= \frac{(t_1-t)}{n} $   and take the limit $ n\rightarrow \infty$  :

\begin{eqnarray}
P(\psi_1,t_1;\psi,t)&=&lim_{n \rightarrow \infty}\langle \psi_1 | (1-\frac{i}{\hbar}(\hat H\epsilon)^n| \psi \rangle \nonumber\\
&=& lim_{n \rightarrow \infty}\langle \psi_1 | (1-\frac{i}{\hbar}(\hat H\epsilon_1)...(1-\frac{i}{\hbar}(\hat H\epsilon_n)| \psi \rangle
\end{eqnarray}

Inserting the resolution of unity

\begin{eqnarray}
\int d\mu_j (\xi) |\psi \rangle \langle \psi |=1
\end{eqnarray}

in the space with fixed S into each time intervals of (7), we can rewrite P as

\begin{eqnarray}
P&=&lim_{n \rightarrow\infty}\sum_j\int...\int(\prod_{k=1}^{n-1} d\mu_j (\psi_k))\prod_{k=1}^{m}\langle \psi_k | (1-\frac{i}{\hbar}(\hat H\epsilon)^n| \psi_{k-1} \rangle \nonumber\\
&=&lim_{n \rightarrow\infty}\sum_j\int...\int(\prod_{k=1}^{n-1} d\mu_j (\psi_k))\prod_{k=1}^{m}\langle \psi_k | \psi_{k-1} \rangle \nonumber\\
& &\times \prod_{k=1}^{n}(1-\frac{i\epsilon}{\hbar}\frac {\langle \psi_k |\hat H | \psi_{k-1} \rangle}{\langle \psi_k | \psi_{k-1} \rangle})
\end{eqnarray}

Where $ \psi_0 =\psi$  and  $ \psi_n =\psi_1$ .

In the limit $ \epsilon\rightarrow0$  the term in the final bracket can be replaced by 

\begin{eqnarray}
(1-\frac{i\epsilon}{\hbar}\frac {\langle \psi_k |\hat H | \psi_{k-1} \rangle}{\langle \psi_k | \psi_{k-1} \rangle})=exp(-\frac{i\epsilon}{\hbar}\frac {\langle \psi_k |\hat H | \psi_{k-1} \rangle}{\langle \psi_k | \psi_{k-1} \rangle})
\end{eqnarray}

Inner product of two coherent states is

\begin{eqnarray}
\langle \psi_k  | \psi_{k-1} \rangle&=&\bar C_0^k  C_0^{k-1}+\bar C_1^k  C_1^{k-1} \nonumber\\
&=& \bar C_0 C_0^{'}+ \bar C_1 C_1^{'} \nonumber\\
&=&1-\frac{\partial}{\partial\theta^{'}}\langle \psi  | \psi^{'} \rangle|_{\theta=\theta^{'}} \triangle\theta-\frac{\partial}{\partial\phi^{'}}\langle \psi  | \psi^{'} \rangle|_{\phi=\phi^{'}} \triangle\phi
\end{eqnarray}

then
\begin{eqnarray}
\langle \psi_k  | \psi_{k-1} \rangle=1+\frac{i}{2}cos\theta \triangle \phi=exp(\frac{icos\theta\triangle\phi}{2})
\end{eqnarray}

The factor $ \prod_{k=1}^{n}\langle \psi_k  | \psi_{k-1} \rangle$   is expressed as 

\begin{eqnarray}
 \prod_{k=1}^{n}\langle \psi_k  | \psi_{k-1} \rangle=exp(\sum_{k=1}^n(\epsilon \frac{1}{\epsilon} ln\langle \psi_k  | \psi_{k-1} \rangle)
\end{eqnarray}

In continuous limit we have

\begin{eqnarray}
exp(\frac{j}{2}\int_t^{t_1}i cos\theta \phi_t d\tau)
\end{eqnarray}

Then we obtain the final expression for the transition amplitude P:
\begin{eqnarray}
P(\psi_1,t_1;\psi,t)&=&lim_{n \rightarrow \infty}\sum_j \int ...\int \prod_{k=1}^{n-1} d\mu_j(\psi) \nonumber\\
& &\times  exp(\frac{i}{2}cos\theta\triangle\phi-\langle \psi |\hat H| \psi\rangle)
\end{eqnarray}

Then in the general form 

\begin{eqnarray}
P(\psi_1,t_1;\psi,t)&=&lim_{n \rightarrow \infty}\sum_j \int D\mu_j(\psi) exp(\frac{-i}{\hbar}\int_t^{t_1}L_j(\theta,\phi)d\tau)
\end{eqnarray}

that

\begin{eqnarray}
L=\frac{j\hbar}{2}cos\theta \phi_t -H_j , &  & {  } H_j(\theta, \phi)=\langle \psi |\hat H| \psi \rangle
\end{eqnarray}

L is Lagrangian. We may rewrite the expression (16) as the $“formal” $ functional integral

\begin{eqnarray}
P&=&\sum_j\int D\mu_j(\psi)exp(-\frac{i}{\hbar}S_j \nonumber\\
S_j&=& \int_t^{t_1}L_j d\tau
\end{eqnarray}

In the case where $\hbar$   is extremely small compared with the action S , the main contribution to the transition amplitude P comes from the path which makes the action stationary with fixed endpoint conditions $ \psi=\psi(t)$  and $\psi_t =\psi(t_1)$  :

\begin{eqnarray}
\delta S&=&0 \nonumber\\
& &\Rightarrow\frac{d}{dt}\frac{\partial L}{\partial \phi_t}-\frac{\partial L}{\partial\phi}=0 \nonumber\\
& &\Rightarrow\frac{d}{dt}\frac{\partial L}{\partial \theta_t}-\frac{\partial L}{\partial\theta}=0
\end{eqnarray}

Using the expression (17) for L, we obtain classical equations in real parameterization:

\begin{eqnarray}
\hbar \phi_t&=& \frac{1}{sin\theta}\frac{\partial H}{\partial\theta} \nonumber\\
\hbar \theta_t&=& \frac{1}{sin\theta}\frac{\partial H}{\partial\phi}
\end{eqnarray}

\section{Properties of SU(3) group}
We consider reference state as $ (1,0,0)^T$ , and general form of coherent state is in the following form:[11]

\begin{eqnarray}
|\psi \rangle &=& D^{\frac{1}{2}}(\theta, \phi)e^{-i\gamma \hat S^z} e^{2ig\hat Q^{xy}}|0\rangle \nonumber\\
&=& C_0|0\rangle +C_1|1\rangle+C_2|2\rangle
\end{eqnarray}

where

\begin{eqnarray}
D^{\frac{1}{2}}(\theta, \phi) &=& e^{-i\phi \hat S^z}e^{-i\theta \hat S^y} \nonumber\\
\hat Q^{xy}&=& \frac{1}{4i}(S^+S^+-S^-S^-) \nonumber\\ 
&=&\frac{i}{2} \left[
\begin{array}{ccc}
0 & 0 &1 \\
0 & 0 & 0 \\
-1 & 0 & 0 
\end{array}
\right] \;
\end{eqnarray}

If we expand exponential terms in coherent state, obtain coefficients:

\begin{eqnarray}
C_0 &=& e^{i\phi}(e^{-i\gamma}sin^2{\theta/2}cosg+e^{i\gamma}cos^2{\theta/2}sing) \nonumber\\
C_1 &=&\frac{ sin{\theta}}{\sqrt{2}} (e^{-i\gamma}cosg-e^{i\gamma}sing) \nonumber\\
C_2 &=& e^{-i\phi}(e^{-i\gamma}cos^2{\theta/2}cosg+e^{i\gamma}sin^2{\theta/2}sing)
\end{eqnarray}

Two angle,  $\theta $ and $\phi$ , define the orientation of the classical spin vector. The angle $\gamma$  is the rotation of the quadrupole moment about the spin vector. The parameter, g, defines change of the spin vector magnitude and that of the quadrupole moment.

Corresponding expression for spin average in SU(3) group is

\begin{eqnarray}
S^+ &=& e^{i\phi}cos2gsin\theta  \nonumber\\
S^- &=& e^{-i\phi} cos2gsin\theta \nonumber\\
S^z &=& cos2g cos\theta 
\end{eqnarray}

In order to obtain the Lagrangian from path integral method acting by the similar way from equation (6) to (17), we obtain

\begin{eqnarray}
\langle \psi_k  | \psi_{k-1} \rangle&=&\bar C_0^k  C_0^{k-1}+\bar C_1^k  C_1^{k-1}+\bar C_2^k  C_2^{k-1} \nonumber\\
&=& \bar C_0 C_0^{'}+ \bar C_1 C_1^{'}+ \bar C_2 C_2^{'} \nonumber\\
&=&1-\frac{\partial}{\partial\theta^{'}}\langle \psi  | \psi^{'} \rangle|_{\theta=\theta^{'}} \triangle\theta-\frac{\partial}{\partial\phi^{'}}\langle \psi  | \psi^{'} \rangle|_{\phi=\phi^{'}} \triangle\phi \nonumber\\
& &-\frac{\partial}{\partial g^{'}}\langle \psi  | \psi^{'} \rangle|_{g=g^{'}} \triangle g-\frac{\partial}{\partial\gamma^{'}}\langle \psi  | \psi^{'} \rangle|_{\gamma=\gamma^{'}} \triangle\gamma \nonumber\\
&  &
\end{eqnarray}

Then the Lagrangian is 
\begin{eqnarray}
L=\hbar cos2g(cos\theta \phi_t +\gamma_t )-H(\phi, \theta, g, \gamma)
\end{eqnarray}

and classical equations of motions are $(\hbar=1)$ :

\begin{eqnarray}
\theta_t&=&- \frac{1}{cos2g sin\theta}(\frac{\partial H}{\partial\phi}-cos\theta \frac{\partial H}{\partial\gamma}) \nonumber\\
 g_t&=& -\frac{1}{2 sin2g}\frac{\partial H}{\partial\gamma} \nonumber\\ 
\phi_t&=& \frac{1}{cos2g sin\theta}\frac{\partial H}{\partial\theta} \nonumber\\
\gamma_t&=& -\frac{1}{2 sin2g}\frac{\partial H}{\partial g}-\frac{cos\theta}{cos2g sin\theta}\frac{\partial H}{\partial\theta} \nonumber\\
\end{eqnarray}

\section{Properties of SU(4) group}

Coherent state in this group is [12,13]

\begin{eqnarray}
|\psi \rangle &=& D^1(\theta, \phi, \gamma)e^{2ig\hat Q^{xy}}e^{-i\beta \hat S^z} e^{-ik\hat F^{xyz}}|0\rangle \nonumber\\
&=& C_0|0\rangle +C_1|1\rangle+C_2|2\rangle+C_3|3\rangle
\end{eqnarray}

where $|0\rangle$   is reference state and 

\begin{eqnarray}
D^1 (\theta,\phi, \gamma)=e^{-i\phi\hat S^z}e^{-i\theta \hat S^y}e^{-i\gamma\hat S^z}
\end{eqnarray}

is Wigner function. Quadrupole moment is

\begin{eqnarray}
\hat Q^{xy} &=& \frac{1}{4\sqrt{3}i}(S^+S^+-S^-S^-) \nonumber\\
&=&\frac{1}{2i} \left[
\begin{array}{cccc}
0 & 0 &1& 0 \\
0 & 0 & 0 &1\\
-1 & 0 & 0 &0\\
0 &-1 & 0& 0
\end{array}
\right] \;
\end{eqnarray}

Octupole moment is 

\begin{eqnarray}
\hat F^{xyz} &=& \frac{1}{6i}(S^+S^+S^+-S^-S^-S^-) \nonumber\\
&=&\frac{1}{i} \left[
\begin{array}{cccc}
0 & 0 &0& 1 \\
0 & 0 & 0 &0\\
0 & 0 & 0 &0\\
-1 &0 & 0& 0
\end{array}
\right] \;
\end{eqnarray}

If we expand exponential term in (29) we obtain the following form:

\begin{eqnarray}
D^1 (\theta ,\phi , \gamma)= \left[
\begin{array}{cccc}
f_1 e^{-\frac{3}{2}i(\phi+\gamma)} & -f_3e^{-\frac{i}{2}(3\phi+\gamma)} &f_4e^{-\frac{i}{2}(3\phi-\gamma)}& -f_2e^{-\frac{3}{2}i(\phi-\gamma)} \\
f_3e^{-\frac{i}{2}(\phi+3\gamma)} & f_5e^{-\frac{i}{2}(\phi+\gamma)} & -f_6e^{-\frac{i}{2}(\phi-\gamma)} &f_4e^{-\frac{i}{2}(\phi-\gamma)}\\
f_4e^{\frac{i}{2}(\phi-3\gamma)} &f_6e^{\frac{i}{2}(\phi-\gamma)} &f_5e^{\frac{i}{2}(\phi+\gamma)} &-f_3e^{\frac{i}{2}(\phi+3\gamma)}\\
f_2e^{\frac{3}{2}i(\phi-\gamma)} &f_4e^{\frac{i}{2}(3\phi-\gamma)} & f_3e^{\frac{i}{2}(3\phi+\gamma)}& f_1e^{\frac{3}{2}i(\phi+\gamma)}
\end{array}
\right] \; \nonumber\\
& &
\end{eqnarray}

where

\begin{eqnarray}
f_1(\theta) &=&cos^3(\theta/2) ,  f_2(\theta)=sin^3(\theta/2), \nonumber\\
f_3(\theta)&=&\sqrt{3}cos^2(\theta/2)sin(\theta/2), f_4(\theta)=\sqrt{3} cos(\theta/2)sin^2(\theta/2) \nonumber\\
f_5(\theta)&=&cos\theta/2(1-3sin^2\theta/2), f_6=sin(\theta/2)(2-3sin^2\theta/2)
\end{eqnarray}

If we insert all above calculation in coherent state (27), obtain:
\begin{eqnarray}
C_0 &=&A_1e^{\frac{3}{2}i(\phi-\gamma-\beta)} -A_2e^{\frac{i}{2}(3\phi+\gamma-3\beta)}+B_1e^{\frac{i}{2}(3\phi-\gamma+3\beta)} \nonumber\\
& &+B_2e^{\frac{3}{2}i(\phi+\gamma+\beta)} \nonumber\\
C_1 &=&A_3e^{\frac{3}{2}i(\phi-\gamma+\beta)} -A_4e^{\frac{i}{2}(\phi+\gamma-3\beta)}+B_3e^{\frac{i}{2}(\phi-\gamma+3\beta)} \nonumber\\
& &-B_4e^{\frac{i}{2}(\phi+3\gamma+3\beta)} \nonumber\\
C_2 &=&B_4^{'}e^{-\frac{i}{2}(\phi+3\gamma+3\beta)} +B_4^{'}e^{\frac{i}{2}(\phi-\gamma+3\beta)}+A_4^{'}e^{-\frac{i}{2}(\phi+\gamma-3\beta)} \nonumber\\
& &-A_2^{'}e^{-\frac{i}{2}(\phi-3\gamma-3\beta)} \nonumber\\
C_3 &=&B_1^{'}e^{-\frac{3}{2}i(\phi+\gamma+\beta)} -B_2^{'}e^{-\frac{i}{2}(3\phi-\gamma+3\beta)}-A_1^{'}e^{-\frac{3}{2}i(\phi-\gamma-\beta)} \nonumber\\
& &+A_2^{'}e^{-\frac{i}{2}(3\phi+\gamma-3\beta)} \nonumber\\
\end{eqnarray}

The coefficients $A_i, B_i,A_i^{'}$   and $ B_i^{'}$ are:
\begin{eqnarray}
A_i &=& a_i sink ,B_i=b_i cosk \nonumber\\
A_i^{'}&=& a_i cosk , B_i^{'}=b_isink
\end{eqnarray}

and
\begin{eqnarray}
a_1&=& sin^3\theta/2 cosg, b_1=\sqrt{3}sin^2\theta/2cos\theta/2 sing \nonumber\\
a_2&=&\sqrt{3}sin\theta/2 cos^2\theta/2sing, b_2=cos^3\theta/2cosg \nonumber\\
a_3&=&\sqrt{3}sin^2\theta/2cos\theta/2cosg, b_3=sin\theta(2-3sin^2\theta/2)sing \nonumber\\
a_4&=&cos\theta(1-sin^2\theta/2)sing , b_4=\sqrt{3}sin\theta/2cos^2\theta/2cosg
\end{eqnarray}

Expression for average spin in SU(4) group is 

\begin{eqnarray}
S^+ &=&\frac{3}{2} e^{i\phi}(1-4cos^2g)cos2ksin\theta  \nonumber\\
S^- &=&\frac{3}{2} e^{-i\phi} (1-4cos^2g)cos2ksin\theta \nonumber\\
S^z &=&\frac{3}{2}(1-4 cos^2g)cos2k cos\theta 
\end{eqnarray}

In similar method that we obtain Lagrangian form path integral for SU(2) and SU(3) groups, for SU(4) group Lagrangian obtain in the following form $(\hbar=1)$ :

\begin{eqnarray}
L=cos2kcos^2g(3cos^2g\beta_t+cos\theta \phi_t+\gamma_t)-H
\end{eqnarray}

Classical equations for motions are:

\begin{eqnarray}
\theta_t&=&\frac{1}{cos2kcos^2g sin\theta}(\frac{\partial H}{\partial\phi}-cos\theta\frac{\partial H}{\partial\gamma}) \nonumber\\
\phi_t&=&-\frac{1}{cos2kcos^2 gsin\theta}\frac{\partial H}{\partial\theta} \nonumber\\
g_t&=&\frac{1}{6cos2kcos^3gsing}\frac{\partial H}{\partial\beta}-\frac{1}{cos2ksin2g}\frac{\partial H}{\partial\gamma} \nonumber\\
\gamma_t&=&\frac{cos\theta}{cos2k cos^2 g sin\theta}\frac{\partial H}{\partial\theta}+\frac{1}{2cos2kcosgsing}\frac{\partial H}{\partial g}+\frac{1}{sin2kcos^2g}\frac{\partial H}{\partial k} \nonumber\\
k_t&=&\frac{1}{sin2kcos^2 g}\frac{\partial H}{\partial\gamma}-\frac{1}{6sin2kcos^4g}\frac{\partial H}{\partial\beta} \nonumber\\
\beta_t&=&-\frac{1}{6sin2kcos^4g}\frac{\partial H}{\partial k}-\frac{1}{6cos2kcos^3 gsing}\frac{\partial H}{\partial g} \nonumber\\
& &
\end{eqnarray}

If we go from SU(4) group to SU(3), we must $g=0,\beta=0$  and in equations $ k\rightarrow g$  in this condition we obtain equations of SU(3) group.

\section{Properties of SU(5) group}

Coherent state in this group is 

\begin{eqnarray}
|\psi \rangle &=& D^{\frac{3}{2}}(\theta, \phi, \gamma)e^{2ig\hat Q^{xy}}e^{-i\beta \hat S^z} e^{-ik\hat O^{xyz}} e^{-im\hat S^z}e^{-in\hat X^{xyzl}}|0\rangle \nonumber\\
&=& C_0|0\rangle +C_1|1\rangle+C_2|2\rangle+C_3|3+C_4|4\rangle
\end{eqnarray}

Where $|0\rangle$   is reference state and 

\begin{eqnarray}
D^{\frac{3}{2}} (\theta,\phi, \gamma)=e^{-i\phi\hat S^z}e^{-i\theta \hat S^y}e^{-i\gamma\hat S^z}
\end{eqnarray}

Quadrupole moment is

\begin{eqnarray}
\hat Q^{xy}=\frac{1}{2i} \left[
\begin{array}{ccccc}
0 & 0 &1& 0&0 \\
0 & 0 & 0 &1&0\\
-1 & 0 & 0 &0&1\\
0 &-1 & 0& 0&0 \\
0&0&-1&0&0
\end{array}
\right] \;
\end{eqnarray}

Octupole moment is 

\begin{eqnarray}
\hat O^{xyz}&=&\frac{1}{12i}(S^+S^+S^+-S^-S^-S^-) \nonumber\\
& &=\frac{1}{i}
 \left[
\begin{array}{ccccc}
0 & 0 &0& 1&0 \\
0 & 0 & 0 &0&1\\
0 & 0 & 0 &0&0\\
-1 &0 & 0& 0&0 \\
0&-1&0&0&0
\end{array}
\right] \;
\end{eqnarray}

Hexadecimalpole moment is 

\begin{eqnarray}
\hat X^{xyzl}&=&\frac{1}{24i}(S^+S^+S^+S^+-S^-S^-S^-S^-) \nonumber\\
& &=\frac{1}{i}
 \left[
\begin{array}{ccccc}
0 & 0 &0& 0&1 \\
0 & 0 & 0 &0&0\\
0 & 0 & 0 &0&0\\
0 &0 & 0& 0&0 \\
-1&0&0&0&0
\end{array}
\right] \;
\end{eqnarray}

If we expand exponential term in (40) we obtain the following form:

\begin{eqnarray}
D^{\frac{3}{2}}(\theta,\phi,\gamma)= 
 \left[
\begin{array}{ccccc}
f_1e^{2i(\gamma+\phi)} & f_2e^{i(2\gamma+\phi)} &f_3e^{2i\gamma}& f_4e^{i(2\gamma-\phi)}&f_5e^{2i(\gamma-\phi)} \\
f_6e^{i(\gamma+2\phi)} &f_7e^{i(\gamma+\phi)} & f_8e^{i\gamma} &f_9e^{i(\gamma-\phi)}&f_4e^{i(\gamma-2\phi)}\\
f_3e^{2i\phi} &-f_8e^{i\phi} & f_{10} &f_8e^{-i\phi}&f_3e^{-2i\phi}\\
-f_4e^{i(-\gamma+2\phi)}&f_9e^{i(-\gamma+\phi)} & -f_8e^{-i\gamma}&f_7e^{-i(\gamma+\phi)}&-f_6e^{-i(\gamma+2\phi)} \\
f_5e^{-2i(\gamma-\phi)}&-f_4e^{i(-2\gamma+\phi)}&f_3e^{-2i\gamma}&-f_2e^{-i(2\gamma+\phi)}&f_1e^{-2i(\gamma+\phi)}
\end{array}
\right] \;\nonumber\\
& &
\end{eqnarray}

where
\begin{eqnarray}
f_1 &=& 1-\frac{\theta^2}{2}+\frac{5\theta^4}{48}-\frac{17\theta^6}{1440}+\frac{13\theta^8}{16128}-\frac{257\theta^{10}}{7257600}+...  \nonumber\\
f_2 &=& -\theta+\frac{5\theta^3}{12}-\frac{17\theta^5}{240}+\frac{13\theta^7}{2016}-\frac{257\theta^9}{725760}+...  \nonumber\\
f_3 &=& \frac{1}{2}\sqrt{\frac{3}{2}}\theta^2-\frac{\theta^4}{2\sqrt{6}}+\frac{\theta^6}{15\sqrt{6}}-\frac{\theta^8}{210\sqrt{6}}+\frac{\theta^{10}}{4725\sqrt{6}}-... \nonumber\\
f_4 &=&-\frac{\theta^3}{4}+\frac{\theta^5}{16}-\frac{\theta^7}{160}+\frac{17\theta^9}{48384}-...   \nonumber\\
f_5 &=&\frac{\theta^4}{16}-\frac{\theta^6}{96}+\frac{\theta^8}{1280}-\frac{17\theta^{10}}{483840}+...   \nonumber\\
f_6 &=&\theta-\frac{5\theta^3}{12}+\frac{17\theta^5}{240}-\frac{13\theta^7}{2016}+\frac{257\theta^9}{725760}-...   \nonumber\\
f_7 &=&1-\frac{5\theta^2}{4}+\frac{17\theta^4}{48}-\frac{13\theta^6}{288}+\frac{257\theta^8}{80640}-\frac{41\theta^{10}}{290304}+...   \nonumber\\
f_8 &=& -\sqrt{\frac{3}{2}}\theta+\sqrt{\frac{2}{3}}\theta^3-\frac{1}{5}\sqrt{\frac{2}{3}}\theta^5+\frac{2}{105}\sqrt{\frac{2}{3}}\theta^7-\frac{1}{945}\sqrt{\frac{2}{3}}\theta^9+...   \nonumber\\
f_9 &=& \frac{3\theta^2}{4}-\frac{5\theta^4}{16}+\frac{7\theta^6}{160}-\frac{17\theta^8}{5376}+\frac{341\theta^{10}}{2419200}-...   \nonumber\\
f_{10} &=& 1-\frac{3\theta^2}{2}+\frac{\theta^4}{2}-\frac{\theta^6}{15}+\frac{\theta^8}{210}-\frac{\theta^{10}}{4725}+...  
\end{eqnarray}

If we insert all above calculation in coherent state (40), obtain:

\begin{eqnarray}
C_0 &=& -e^{2i(\beta+m)}sinn(Ae^{2i(\phi-\gamma)}f_5+Be^{2i(\phi+\gamma)}f_1+Ce^{2i\phi}f_3) \nonumber\\
&  &+cosne^{-2im} (e^{i\beta}(cosge^{i(2\phi-\gamma)}f_4+singe^{i(2\phi+\gamma)}f_2)sink \nonumber\\
&   & +e^{-2i\beta}(Be^{2i(\phi-\gamma)}f_5+Ae^{2i(\phi+\gamma)}f_1-Ce^{2i\phi}f_3)cosk ) \nonumber\\
C_1&=& -e^{2i(\beta+m)}sinn(Ae^{i(\phi-2\gamma)}f_4+Be^{i(\phi+2\gamma)}f_6+Ce^{i\phi}f_8) \nonumber\\
&  &+cosne^{-2im} (e^{i\beta}(cosge^{i(\phi-\gamma)}f_9+singe^{i(\phi+\gamma)}f_7)sink \nonumber\\
&   & +e^{-2i\beta}(Be^{i(\phi-2\gamma)}f_4+Ae^{i(\phi+2\gamma)}f_6-Ce^{i\phi}f_8)cosk ) \nonumber\\
C_2 &=& -e^{2i(\beta+m)}sinn(Ae^{-2i\gamma}f_3+Be^{2i\gamma}f_3+Cf_{10}) \nonumber\\
&  &+cosne^{-2im} (e^{i\beta}(cosge^{-i\gamma}f_8-singe^{i\gamma}f_8)sink \nonumber\\
&   & +e^{-2i\beta}(Be^{-2i\gamma}f_3+Ae^{2i\gamma}f_3-Cf_{10})cosk ) \nonumber\\
C_3 &=& e^{2i(\beta+m)}sinn(Ae^{-i(\phi+2\gamma)}f_6+Be^{i(-\phi+2\gamma)}f_4+Ce^{-i\phi}f_8) \nonumber\\
&  &+cosne^{-2im} (e^{i\beta}(cosge^{-i(\phi+\gamma)}f_7+singe^{i(-\phi+\gamma)}f_9)sink \nonumber\\
&   & -e^{-2i\beta}(Be^{-i(\phi+2\gamma)}f_6+Ae^{i(-\phi+2\gamma)}f_4-Ce^{-i\phi}f_8)cosk ) \nonumber\\
C_4 &=& -e^{2i(\beta+m)}sinn(Ae^{-2i(\phi+\gamma)}f_1+Be^{2i(-\phi+\gamma)}f_5+Ce^{-2i\phi}f) \nonumber\\
&  &+cosne^{-2im} (e^{i\beta}(-cosge^{-i(2\phi+\gamma)}f_2-singe^{i(-2\phi+\gamma)}f_4)sink \nonumber\\
&   & +e^{-2i\beta}(Be^{-2i(\phi+\gamma)}f_1+Ae^{2i(-\phi+\gamma)}f_5-Ce^{-2i\phi}f_3)cosk ) \nonumber\\
\end{eqnarray}

where
\begin{eqnarray}
A=\frac{1}{2}(1+cos(\sqrt{2}g)), B=\frac{1}{2}(1-cos(\sqrt{2}g)), C=\frac{sin(\sqrt{2}g)}{\sqrt{2}}
\end{eqnarray}

Expression for average spin in SU(5) group is 

\begin{eqnarray}
S^+ &=&2 e^{i\phi}cos(\sqrt{2}g)(1-4cos^2k)cos2n sin\theta  \nonumber\\
S^- &=&2 e^{-i\phi}cos(\sqrt{2}g)(1-4cos^2k)cos2n sin\theta  \nonumber\\
S^z &=&2cos(\sqrt{2}g)(1-4cos^2k)cos2n sin\theta  \nonumber\\ 
\end{eqnarray}

Lagrangian in this group is

\begin{eqnarray}
L=2\hbar cos2n cos^2 k cos^2 g (3cos^2 k cos^2 g \beta_t+3 cos^2 k m_t +cos\theta \phi_t +\gamma_ t)-H
\end{eqnarray}

Classical equations for motion are:

\begin{eqnarray}
\theta_t&=&\frac{1}{cos2ncos^2g sin\theta cos^2k}\frac{\partial H}{\partial\phi}-\frac{cos\theta}{cos2n cos^2g cos^2k sin\theta}\frac{\partial H}{\partial\gamma} \nonumber\\
\phi_t&=&-\frac{1}{cos2ncos^2 gcos^2ksin\theta}\frac{\partial H}{\partial\theta} \nonumber\\
g_t&=&\frac{1}{6cos2ncos^3gsingcos^4 k}\frac{\partial H}{\partial\beta}-\frac{1}{3cos2nsin2gcos^4 k}\frac{\partial H}{\partial m} \nonumber\\
\gamma_t&=&\frac{cos\theta}{cos2n cos^2 g sin\theta cos^2k}\frac{\partial H}{\partial\theta}+\frac{1}{2cos2ncos^2gsink cosk}\frac{\partial H}{\partial k} \nonumber\\
& &-\frac{1}{sin2ncos^2g cos^2k}\frac{\partial H}{\partial n} \nonumber\\
k_t&=&\frac{1}{6sinkcos^2 gcos2ncos^3k}\frac{\partial H}{\partial m}-\frac{1}{2sinkcos^2gcosk cos2n}\frac{\partial H}{\partial\gamma} \nonumber\\
\beta_t&=&\frac{1}{6sin2ncos^4gcos^4k}\frac{\partial H}{\partial n}-\frac{1}{6cos2ncos^3 gsingcos^4 k}\frac{\partial H}{\partial g} \nonumber\\
& &-\frac{1}{sin2ncos^2g cos^2k}\frac{\partial H}{\partial n} \nonumber\\
n_t&=&\frac{1}{cos2ncos^2gcos^2 k sin\theta}\frac{\partial H}{\partial\gamma}-\frac{1}{6sin2ncos^4gcos^4k}\frac{\partial H}{\partial\beta} \nonumber\\
m_t&=& -\frac{1}{cosgcos^4 ksin2n}\frac{\partial H}{\partial g}-\frac{1}{cos^2g cos^3k cos2nsink}\frac{\partial H}{\partial k}   \nonumber\\
& &
\end{eqnarray}

If in equation (51) we insert $ g=0,m=0, n\rightarrow k, k\rightarrow g$  obtain equation (39) that is equations in SU(4) group.

\section{Discussion}
Magnetically ordered materials (magnets) are known as essentially nonlinear systems. Localized nonlinear excitations with finite energy, or solitons, play an important role in description of nonlinear dynamics, in particular, spin dynamics for low-dimensional magnets, with different kind of magnetic order. To date, solitons in Heisenberg ferromagnets, whose dynamics are described by the Landau-Lifshitz equation for the constant-length magnetization vector, have been studied in details. In terms of microscopic spin models, this picture corresponds to the exchange Heisenberg Hamiltonian, with the isotropic bilinear spin interaction $ J\vec S_1.\vec S_2$. For a spin of  $S > 1/2$ the isotropic interaction is not limited by this term and can include higher invariants such as $( J_n\vec S_1.\vec S_2)^n$ with n up to 2S.  In the case of exchange anisotropy only,  Landau-Lifshitz equation can fully describe spin dynamics of Heisenberg magnets. If we take into account single-ion anisotropy, 
 Landau-Lifshitz equation can be used only for  $S=\frac{1}{2}$ Heisenberg Ferromagnet, in the case of higher spin $S\geq \frac{1}{2}$  excitations of multipole spin dynamics should be involved in the model in order to provide full description of the spin system.  Suggested generalized coherent states, which are constructed on the case of appropriate group, commonly say SU(2S+1), for the system S systems are able to provide necessary number of variables  of classical spin (multipole) dynamics.


\begin{thebibliography}{9}
\bibitem{law}	R. J. Glauber. Phys. Rev. 130, 2529 (1963).	
\bibitem{law}	A. M. Perelomov. Generalized coherent states and applications. Nauka, Mascow, 1987.
\bibitem{law}	R. Gilmore. Ann. Phys., 74 391 (1972).
\bibitem{law}	W.-M. Zhang, D. H. Feng and R. Gilmore. Rev. Mod. Phys. 62, 867 (1990).
\bibitem{law}	F. W. Hehi, Yu. N. Obukhov, J.-P. Rivera and H. Schmid, Phys. Rev. A, 77, 0221106 (2008).
\bibitem{law}	D. N. Astrov, N. B. Ermakov, and S. V. Korostin, JETP Letters, 67, No. 1, 15-20 (1998).
\bibitem{law}	D. N. Astrov, N. B. Ermakov, and S. V. Korostin. Pis’ma v ZhETF, JETP Letters 59, 274 (1994) (in Russian).
\bibitem{law}	I. E. Dzyaloshinskii. Sol, St. Comm. 82, 579 (1992).
\bibitem{law}	H. Kuratsuji and T. Suzuki. J.Math. Phys. 21(3) 1980.
\bibitem{law}	J. J. Sakurai. Modern quantum mechanics, 1999.
\bibitem{law}	V.S.Ostrovskii. Sov. Phys. JETP 64(5), 999, 1986.
\bibitem{law}	Kh. O. Abdulloev, Kh. Kh. Muminov. Coherent states of SU(4) group in real parameterization and Hamiltonian equations of motion. Reports of Tajikistan Academy of science V.36, N6, I993 (in Russian).
\bibitem{law}	Kh. O. Abdulloev, Kh. Kh. Muminov. Accounting of quadrupole dynamics of magnets with spin  . Proceedings of Tajikistan Academy of Sciences, N.1, 1994, P.P. 28-30 (in Russian).
\end{thebibliography}
\end{document}